\documentstyle[amssymb,11pt]{article}

\input{tcilatex}
\begin{document}

\title{The field state dissipative dynamics of two-photon Jaynes-Cummings model
with Stark shift in dispersive approximation}
\author{L. Zhou\thanks{%
E-mail:zhlhxn@dlut.edu.cn}, H. S. Song\thanks{%
E-mail:hssong@dlut.edu.cn}, Y. X. Luo \\
Department of Phys., Dalian University of Technology, Dalian 116023, China.}
\maketitle

\vspace{0.1in}\baselineskip=17.8pt

\noindent \rule{\textwidth}{0.5pt}%

{\bf Abstract:}

We present the field state dissipative dynamics of two-photon
Jaynes-Cummings model (JCM) with Stark shift in dispersive approximation and
investigate the influence of dissipation on entanglement. We show the
coherence properties of the field is also affected by the cavity when
nonlinear two-photon process is involved


\noindent keyword: the field state dissipative evolution, two-photon JCM,
the coherence loss.

\noindent \rule{\textwidth}{0.5pt}%

\section{Introduction}

Jaynes-Cummings Model (JCM) [1] has been recognized as the simplest and most
effective model to describes a two-level atom interaction with the
electromagnetic field. It presents an extremely rich and nontrivial
dynamics. In addition to its exact solvability within the rotating-wave
approximation, the most interesting aspect of its dynamics is the
entanglement between the atom and the field. Much attention has been focused
on the entanglement of the field and the atom in JCM[2-6]. Recently,
entanglement as a physical resource has been used in quantum information
science such as quantum teleportation [7], superdense coding [8] and quantum
cryptography [9]. In spite of the success achieved by quantum theory in what
concerns the prediction of experiments in general, there has been a lot of
debate about some of its fundamental aspects, one of which is directly
related to the entanglement, which presents its most famous illustrations in
Einstein-Podolsky-Rosen paradox. People have gain some satisfactory answer
about Schr\"{o}dinger's cat paradox and why the entanglement phenomenon does
not occur in the classical world. Among those the one that stress the role
played by the environment, which is represented by a thermal reservoir.
Decoherence due to the irreversible coupling of the observed system to the
outside world reservoir eventually turn correlated state into a statistical
mixture.

The details of the entanglement between two subsystem in the presence of
such an environment is worth to study. In Ref. [10] dispersive atomic
evolution in a dissipative- driven cavity were studied. The influence of
driving field on the quantized driven field and on atom properties in both
the dissipative and the lossless cases were obtained. And in Ref.[11], the
authors employed JCM in the dispersive approximation in a dissipative cavity
at zero temperature to study the entanglement between the atom and the field
as well as the decoherence induced by the cavity. They had shown the cavity
has practically no influence in the coherence properties of the field from
the qualitative point of view but the atom's coherence properties are
strongly influenced by dissipation both qualitatively and quantitatively,
although it is not directly coupled to the cavity. The purpose of this paper
is to research the dynamics of a two-level atom with Stark shift interaction
with the field by two-photon process in a dissipative cavity and to study
whether the cavity has practically no influence in the coherence properties
of the field when two-photon process is involved. We also plane to research
the entanglement influenced by the dissipation. We show that the coherence
properties of the field is also affected by the cavity when nonlinear
two-photon process is involved. We also observe the entanglement is
influenced by dissipation and make the amplitude of oscillation suppress. In
two-photon process the relation the coherence loss of the field with the
intensity of the field is also given.

\section{\protect\smallskip Two-photon Jaynes-Cummings Model with Stark
shift in dispersive approximation}

\bigskip The Hamiltonian of the two-photon JCM including Stark shift with
rotating-wave approximation [4,12] is given by

\begin{equation}
\hat{H}=\omega \hat{a}^{+}\hat{a}+\frac{\omega _0}2\hat{S}_z+\hat{a}^{+}\hat{%
a}(\beta _2|e\rangle \langle e|+\beta _1|g\rangle \langle g|)+\lambda (\hat{a%
}^{+2}\hat{S}_{-}+\hat{a}^2\hat{S}_{+}),
\end{equation}
where $\omega $ is the field frequency, $\omega _0$ is the frequency between
the two-level ( denoted by $e$ and $g$ ) of the atom , represented here by
the well-know Pauli matrices $\hat{S}_i$, $\beta _1$ and $\beta _2$ are
effective Stark shift coefficients, which related to $\lambda _1$ and $%
\lambda _2$ and $\Delta $ (detuning) as follows:

\begin{equation}
\beta _i=\frac{\lambda _i^2}\Delta ,i=1,2;\lambda =\frac{\lambda _1\lambda _2%
}\Delta ,
\end{equation}
where $\lambda _1$ and $\lambda _2$ denote intermediate state $|i\rangle $
coupling to $|e\rangle $ and $|g\rangle $ with strengths, $\lambda $ measure
the two-photon atom-field coupling. In an invariant subspace spanned by $%
|e\rangle \otimes |n\rangle $ and $|g\rangle \otimes |n+2\rangle ,$the
Hamiltonian takes the following form

\begin{equation}
\hat{H}=\left[ 
\begin{array}{ll}
\omega n+\frac{\omega _0}2+\beta _2n & \lambda \sqrt{(n+1)(n+2)} \\ 
\lambda \sqrt{(n+1)(n+2)} & \omega n+\frac{\omega _0}2+\delta +\beta _1(n+2)
\end{array}
\right] ,
\end{equation}
where detuning $\delta =2\omega -\omega _0$ measures how off resonance the
two system are. The eigenvalues are

\begin{eqnarray}
E_{\pm } &=&\omega n+\frac{\omega _0}2+\frac{\beta _1(n+2)}2+\frac{\beta _2n}%
2-\frac \delta 2  \nonumber \\
&&\pm \frac 12\{[\beta _1(n+2)-\beta _2(n+1)+\beta _2-\delta ]^2+4\lambda
^2(n+1)(n+2)\}^{\frac 12}.
\end{eqnarray}
The dispersive limit of two-photon process with Stark shift is obtained when 
$\hat{H}_I$ can be considered as a small perturbation in the following sense:

\begin{equation}
\frac{\beta _1(n+2)}{|\delta -\beta _2|}\ll 1,\frac{\beta _2(n+1)}{|\delta
-\beta _2|}\ll 1\text{ }
\end{equation}
for any relevant $n.$

\begin{eqnarray}
E_{+} &=&\omega n+\frac{\omega _0}2+\beta _2n+\frac{\beta _1\beta _2}{%
|\delta -\beta _2|}(n+1)(n+2),  \nonumber \\
E_{-} &=&\omega n+\frac{\omega _0}2-\delta +\beta _1(n+2)-\frac{\beta
_1\beta _2}{|\delta -\beta _2|}(n+1)(n+2).
\end{eqnarray}

If the condition Eq.(5) is fulfilled for all $n$ value, we can work with the
effective Hamiltonian 
\begin{eqnarray}
\hat{H}_{eff} &=&\omega \hat{a}^{+}\hat{a}+\frac{\omega _0}2\hat{S}_z+\hat{a}%
^{+}\hat{a}(\beta _2|e\rangle \langle e|+\beta _1|g\rangle \langle g|) 
\nonumber \\
&&+\Omega [(\hat{a}^{+}\hat{a}+1)(\hat{a}^{+}\hat{a}+2)|e\rangle \langle e|-%
\hat{a}^{+}\hat{a}(\hat{a}^{+}\hat{a}-1)|g\rangle \langle g|],
\end{eqnarray}
where for simplicity we let $\frac{\beta _1\beta _2}{\delta -\beta _2}%
=\Omega .$

\section{\protect\bigskip Time evolution of the initial field state}

\bigskip We assume that there is a reservoir coupled to the field in the
usual way. In the dispersive approximation, a two-level atom interacting
with a quantum field in a dissipative cavity has a standard form ($\hslash
=1)$ 
\begin{equation}
\frac{d\hat{\rho}}{dt}=-i[\hat{H}_{eff},\hat{\rho}]+{\cal L}\hat{\rho}.
\end{equation}
The losses in the cavity are phenomenologically represented by the
superoperator ${\cal L}$. At the zero temperature, we have 
\begin{equation}
{\cal L}\hat{\rho}=\kappa (2\hat{a}\hat{\rho}\hat{a}^{+}-\hat{a}^{+}\hat{a}%
\hat{\rho}-\hat{\rho}\hat{a}^{+}\hat{a}),
\end{equation}
where $\kappa $ is the damping constant. In the interaction picture, the
master equation takes the form 
\begin{eqnarray}
\frac{d\hat{\rho}}{dt} &=&-i[\hat{a}^{+}\hat{a}(\beta _2|e\rangle \langle
e|+\beta _1|g\rangle \langle g|)+\Omega [(\hat{a}^{+}\hat{a}+1)(\hat{a}^{+}%
\hat{a}+2)|e\rangle \langle e|  \nonumber \\
&&-\hat{a}^{+}\hat{a}(\hat{a}^{+}\hat{a}-1)|g\rangle \langle g|,\hat{\rho}%
]+\kappa (2\hat{a}\hat{\rho}\hat{a}^{+}-\hat{a}^{+}\hat{a}\hat{\rho}-\hat{%
\rho}\hat{a}^{+}\hat{a}).  \label{p2}
\end{eqnarray}
Two-photon process is a nonlinear process. Completely solving equation (10)
is not a easy task. Here we just plan to consider the subsystem field
property. We write the reduced field operator as

\begin{eqnarray}
\hat{\rho}_{_F}(t) &=&Tr_{_A}\hat{\rho}(t)  \nonumber \\
&=&\hat{\rho}_{_{gg}}(t)+\hat{\rho}_{ee}(t).
\end{eqnarray}
Liouvillians corresponding to the matrix elements $\hat{\rho}_{_{gg}}$ and $%
\hat{\rho}_{ee}$ have the form

\begin{equation}
{\cal L}_{gg}=2\kappa {\cal \digamma }+i\Omega ({\cal M}^2-{\cal P}%
^2)-(\kappa +i\Omega _g){\cal M}-(\kappa -i\Omega _g){\cal P},  \label{L1}
\end{equation}
\begin{equation}
{\cal L}_{ee}=2\kappa {\cal \digamma }-i\Omega ({\cal M}^2-{\cal P}%
^2)-(\kappa +i\Omega _e){\cal M}-(\kappa -i\Omega _e){\cal P},  \label{L2}
\end{equation}
where $\Omega _g=\beta _1+\Omega ,\Omega _e=\beta _2+3\Omega .$ The
superoperators in Eq.(12 ) and (13) are defined as ${\cal \digamma }\hat{\rho%
}=\hat{a}\hat{\rho}\hat{a}^{+},{\cal M}\hat{\rho}=\hat{a}^{+}\hat{a}\hat{\rho%
},{\cal P}\hat{\rho}=\hat{\rho}\hat{a}^{+}\hat{a}$. They satisfy the
commutation relation [10] 
\begin{equation}
\lbrack {\cal \digamma },{\cal M}]={\cal \digamma },\qquad [{\cal \digamma },%
{\cal P}]={\cal \digamma },\qquad [{\cal M},{\cal P}]=0.
\end{equation}
Hence ,the master equation can be solved by applying the dynamical symmetry
method proposed in Ref. [13] and we have

\begin{eqnarray}
\hat{\rho}_{_{gg}}(t) &=&e^{{\cal L}_{gg}t}\hat{\rho}_{_{gg}}(0)  \nonumber
\\
&=&\exp [i\Omega t({\cal M}^2-{\cal P}^2)-(\kappa +i\Omega _g)t{\cal M}%
-(\kappa -i\Omega _g)t{\cal P}]  \nonumber \\
&&\times \exp [(1-e^{-2\kappa t+2i\Omega t({\cal M-P)}})\frac{\kappa {\cal %
\digamma }}{\kappa -i\Omega ({\cal M-P})}]\hat{\rho}_{_{gg}}(0),  \label{PG}
\end{eqnarray}
\begin{eqnarray}
\hat{\rho}_{_{ee}}(t) &=&e^{{\cal L}_{ee}t}\hat{\rho}_{_{ee}}(0)  \nonumber
\\
&=&\exp [-i\Omega t({\cal M}^2-{\cal P}^2)-(\kappa +i\Omega _e)t{\cal M}%
-(\kappa -i\Omega _e)t{\cal P}]  \nonumber \\
&&\times \exp [(1-e^{-2\kappa t-2i\Omega t({\cal M-P)}})\frac{\kappa {\cal %
\digamma }}{\kappa +i\Omega ({\cal M-P})}]\hat{\rho}_{_{ee}}(0),  \label{PE}
\end{eqnarray}

We assume the initial state of the system as

\begin{equation}
|\Psi _{a-f}\rangle =\frac 1{\sqrt{2}}(|e\rangle +|g\rangle )\otimes |\alpha
\rangle ,
\end{equation}
where, as is usual in recent experiments [14], the atom enters the cavity in
a coherence superposition and finds there a coherent field state $|\alpha
\rangle $ , therefore initially $\hat{\rho}_{_{gg}}(0)=\hat{\rho}%
_{ee}(0)=\frac 12|\alpha \rangle \langle \alpha |$, finally we get 
\begin{equation}
\hat{\rho}_{_{gg}}(t){\normalsize =}\frac 12\exp ({\normalsize -|\alpha |}%
^2)\sum_{m,n}\frac{\alpha ^m\alpha ^{*}{}^n}{\sqrt{m!n!}}\exp {\normalsize %
[\Gamma }_{mn}(t)+{\normalsize i\Theta }_{gmn}(t)]|m\rangle \langle n|,
\end{equation}
\begin{equation}
\hat{\rho}_{_{ee}}(t)=\frac 12\exp ({\normalsize -|\alpha |}^2)\sum_{m,n}%
\frac{\alpha ^m\alpha ^{*}{}^n}{\sqrt{m!n!}}\exp [{\normalsize \Gamma }%
_{mn}(t)+{\normalsize i\Theta }_{emn}(t)]|m\rangle \langle n|,
\end{equation}
where 
\begin{eqnarray}
\Gamma _{mn}(t) &=&-\kappa (m+n)t+\frac{|\alpha |^2\kappa }{\kappa ^2+\Omega
^2(m-n)^2}  \nonumber \\
&&\times \{\kappa -e^{-2\kappa t}[\kappa \cos 2\Omega t(m-n)-\Omega
(m-n)\sin 2\Omega t(m-n)]\},
\end{eqnarray}
and 
\begin{eqnarray}
\Theta _{imn}(t) &=&-\Omega _i(m-n)t\pm \Omega (m^2-n^2)t  \nonumber \\
&&\pm \frac{|\alpha |^2\kappa \Omega (m-n)}{\kappa ^2+\Omega ^2(m-n)^2}[%
1-e^{-2\kappa t}\cos 2\Omega t(m-n)],
\end{eqnarray}
where $i=e,g$, when $i=g$ last equation chose $+$ and $-$ corresponding to $%
i=e$. We obtained the $kth$ moment of amplitude 
\begin{eqnarray}
&<&a^n>=\frac 12\alpha ^n\exp {\normalsize [}\Gamma _{n0}(t)+|\alpha
|^2(e^{-2\kappa t}\cos 2\Omega nt-1)]  \nonumber \\
&&\times \{\exp i[|\alpha |^2e^{-2\kappa t}\sin 2\Omega nt+\Theta
_{gn0}(t)]+\exp [i(-|\alpha |^2e^{-2\kappa t}\sin 2\Omega nt+\Theta
_{en0}(t)]\}.
\end{eqnarray}

One may check the decay behavior of the system by measuring the $kth$ moment
of amplitude. Here we aim to discuss the coherence loss of the field by
means of idempotency defect or linear entropy, which is convenient way to
study the coherence properties of the density operator as a function of
time. Linear entropy is defined as [15] 
\begin{equation}
S_f=1-Tr(\hat{\rho}_f^2).  \label{s}
\end{equation}
The quantity $Tr(\hat{\rho}_f^2)$ can be take as a measure of the degree of
purity of the reduced state; for a pure state $S_f$ is zero but for $%
S_f\simeq 1$ the state corresponds to a mixture, with information
effectively lost. From Eq.(15) through Eq. (23), we have 
\begin{equation}
S_f=1-\exp ({\normalsize -2|\alpha |}^2)\sum_{m,n}\frac{|\alpha |^{2(m+n)}}{%
m!n!}\exp [2\Gamma _{mn}(t)]\cos ^2\frac{\Theta _{gmn}(t)-\Theta _{emn}(t)}2.
\label{S}
\end{equation}

The function $\Gamma _{mn}(t)$ in Eq. (20) embody the effect of reservoir
because it vanishes for $k\rightarrow 0$ . Although the complicate
expression in Eq. (24) is not analytical , it contain the function $\Gamma
_{mn}(t)$ , the coherence property of field should be affected by cavity. We
will discuss it in next section.

\section{The property of the field state dissipation}

The function $\Gamma _{mn}(t)$ in Eq. (24) presented in exponential factors
controls the coherence loss of the field. It is always nonpositive and
decrease with time and have the similar form to the usual dissipative JCM
[11]. But the situation is not the same. There the $\Gamma (t)$ function did
not appear in the idempotent defect of the field. On other hand, in Ref.
[11] the field and the atom disentangle at instant $t_d=\frac{n\pi }\omega $
, during disentangle only the field is found in a pure state and $%
s_f(t_d)=0. $ That circumstance is the same as JCM without dissipation ,
thus the coherence properties of the field have no influence by the cavity
in qualitatively. Here due to the two-photon process and the Stark shift the
function $\Gamma _{mn}(t)$ appear in Eq.(24). Except the initial time at no
instants the field is in pure state, because during evolution the value of
linear entropy $s_f(t)\neq 0$ ( Of course, the equilibrium state of the
field corresponds to vacuum, the linear entropy of the field is zero).
Therefore we can reckon the field is also influenced by the cavity in
qualitatively. We will furth show the judgment in Fig.(1).

We plot the field's idempotency defect as a function of time for several $%
\kappa $ values in Fig. (1). It is noticed the behavior of $s_f(t)$ is
complicated, presenting local maxima and minima in wave packet trajectory.
The local maxima and minima is due to the field interaction with a atom,
corresponding to entanglement and disentanglement. Because of the influence
of dissipation on entanglement, the amplitude of local maxima and minima
decrease with time. It is exactly as usual JC with dissipation. Due to the
repeated period of entanglement and disentanglement the state of the atom
and field loss and gain coherence but the coherence recovered by the atom is
never that which was lost. The field finally change into pure state (vacuum
state) and its coherence lost completely. From Fig. (1) we also observe that
the wave packet trajectory is determined by the cavity dissipative. The
larger values of $\kappa $ the more rapid is the field's idempotency defect
reach its asymptotic value zero. We can assure that the coherence property
of the field is affected by cavity in quantitatively.

The dependence of the idempotency defects of the field with the intensity of
the field is shown in Fig. (2). We choose the same value $\kappa $ and
obtain the similar wave packet trajectory of linear entropy. With the
increase of the intensity, the classicality of the field become obvious and
the entanglement between the two subsystem change weaker. The amplitude of
local maxima and minima is suppressed much more. Another contribution of
intensity is to increase the maxima value of $S_f$. Due to enhance of
intensity, the degree of maxima mixture state fortify. We also notice that
with the increase of the intensity of the field, nonlinear behavior of the
field lost and gain its coherence become obvious. In our calculation, we
include the Stark shift but we find that the field coherence loss is
affected little by different values of Stark shift coefficients $\beta _1$
and $\beta _2.$

\section{Conclusion}

We study the dynamics of a two-level atom with Stark shift interaction with
the field by two-photon process in a dissipative cavity and solve the
complicated Liouvil equation. We obtain idempotency defects of the field and
show as follow: (1) the coherence property of the field is affected by
cavity not only in qualitatively but also in quantitatively when two-photon
process and Stark shift is involved. (2) The influence of dissipation on
entanglement make the amplitude of each state suppressed. (3) The larger the
intensity of the field is the weaker the entanglement of subsystem and the
larger maxima degree of mixture state.

\medskip {\small Fig. 1 Idempotency defects of the field as a function of }$%
\Omega t${\small \ for different values of dissipation constant }$\kappa $%
{\small \ . Where }$\alpha =1${\small , }$(\beta _2-\beta _1)/\Omega =0.02$%
{\small \ and (a):}$\kappa /\Omega =0.02${\small ; (b): }$\kappa /\Omega
=0.04${\small ; (c): }$\kappa /\Omega =0.1.$

{\small Fig.2 Idempotency defects of the field as a function of }$\Omega t$%
{\small \ for different values of the intensity of the field. For all plots,
we chose }$\kappa /\Omega =0.04${\small , (}$\beta _2-\beta _1)/\Omega =0.02$%
{\small , where (a):}$\overline{n}=1.0${\small ; (b): }$\overline{n}=2.0$%
{\small ; (c): }$\overline{n}=3.0$

\end{document}